\documentclass[reprint,aip,amsmath,amssymb,nofootinbib]{revtex4-2}
\usepackage{amssymb}
\usepackage{graphicx}
\usepackage{graphics}
\usepackage{amsmath}
\usepackage{placeins}
\usepackage{hyperref}
\usepackage[dvipsnames]{xcolor}
\usepackage{wasysym}
\usepackage{soul}
\usepackage{float}
\usepackage{array}
\usepackage{tabulary}
\usepackage{fancyhdr}
\usepackage{wrapfig}

\usepackage[normalem]{ulem}
\usepackage{graphicx}
\usepackage{color,soul}
\usepackage{subfigure}
\usepackage{dcolumn}
\usepackage{bm}
\usepackage{hyperref}
\usepackage{url}
\usepackage{multirow}
\hypersetup{
    colorlinks=true,
    linkcolor=blue,
    filecolor=blue,      
    urlcolor=blue,
    citecolor=blue,
}
\usepackage{float}

\usepackage[makeroom]{cancel}
\begin{document}

\title{Confirmation of Transit-Time Limited Field Emission in Advanced Carbon Materials with Fast Pattern Recognition Algorithm}

    \author{\firstname{Taha Y.} \surname{Posos}}
	\email{posostah@msu.edu}
    \affiliation{Department of Electrical and Computer Engineering, Michigan State University, MI 48824, USA}
    \author{\firstname{Oksana} \surname{Chubenko}}
    \affiliation{Department of Physics, Arizona State University, AZ 85287, USA}
	\author{\firstname{Sergey V.} \surname{Baryshev}}
	\email{serbar@msu.edu}
	\affiliation{Department of Electrical and Computer Engineering, Michigan State University, MI 48824, USA}

\begin{abstract}
An accurate estimation of the experimental field-emission area remains a great challenge in vacuum electronics. The lack of convenient means, which can be used to measure this parameter, creates a critical knowledge gap, making it impossible to compare theory to experiment. In this work, a fast pattern recognition algorithm was developed to complement a field emission microscopy, together creating a methodology to obtain and analyze electron emission micrographs in order to quantitatively estimate the field emission area. The algorithm is easy to use and made available to the community as a freeware, and therefore is described in detail. Three examples of dc emission are given to demonstrate the applicability of this algorithm to determine spatial distribution of emitters, calculate emission area, and finally obtain experimental current density as a function of the electric field for two technologically important field emitter materials, namely an ultrananocrystalline diamond film and a carbon nanotube fiber. Unambiguous results, demonstrating the current density saturation and once again proving that conventional Fowler-Nordheim theory, its Murphy-Good extension, and the vacuum space-charge effect fail to describe such behaviour, are presented and discussed. We also show that the transit-time limited charge resupply captures the current density saturation behaviour observed in experiments and provides good quantitative agreement with experimental data for all cases studied in this work.
\end{abstract}

\maketitle

\section{Introduction}\label{S:intro}
Many studies\cite{locally,mypaper,Jiahang1,Jiahang2,cui1,kolosko} have convincingly demonstrated that the electron emission from a large surface area field emission cathode placed in a macroscopic electric field $E$ is not uniform. The emission area is only a small portion of the total surface area of a cathode since most of the emission is confined to a small number of emission spots randomly distributed over the cathode surface. Therefore, a proper and thorough estimation of the apparent emission area and a number of emission locations is essential to quantify field emitters in terms of the current density $j$ and its variation with the applied electric field $E$. The importance of developing such methodologies is twofold. First is practical: an actual emission area needs to be known to compare cathode materials produced by various or varied syntheses. Second is fundamental: only a properly established $j$--$E$ (and not $I$--$V$) relationship can be used to define the validity range of a classical Fowler-Nordheim (FN) emission and clarify the role of other mechanisms that could cause deviation from the FN emission, i.e. cause non-conventional behavior that is being observed across a large body of experimental work\cite{locally,mypaper,Jiahang1,Jiahang2,cui1,kolosko,serbun2013,GaN,KOCK,Obraz}.

A simple and convenient way to measure distribution of electron emission sites is using a luminescence (or phosphor) screen, also known as a scintillator. These screens emit light when interact with electrons. When such a screen is used as an anode in a field emission experiment, it magnifies and projects an emission pattern formed on the cathode surface under the external field force. When captured by a camera, it creates a micrograph. An experimental system can be designed such that $I$--$V$ curves can be taken synchronously with micrographs. Then, the micrographs can be used to extract the apparent emission area and obtain the $j-E$ relation. This work is motivated by the lack of a guided micrograph processing for emission area calculations. Here, we present a fast image processing algorithm and demonstrate its application for a thorough analysis of field emission data. Results are presented for micrographs obtained from ultra-nano-crystalline dimond (UNCD) films and carbon nano-tubes (CNTs) under the applied dc field using screens made of Ce-doped yttrium aluminum garnet (YAG:Ce), which produces a bright green luminescence line at 550 nm. The proposed image processing algorithm is also applicable to any other phosphor screen that is able to produce spatially separated features that are bright enough.

The algorithm is realized on the MATLAB platform and takes an advantage of its strength in processing arrayed data. When compared to an earlier version implemented in Mathematica,\cite{locally} where a server was required to process extensive sets of micrographs, our present implementation performs 10 times faster on a personal laptop. The image processing code is an open-source software: the first and future releases as well as examples and the user manual can be found in our GitHub page (see Ref.~\onlinecite{GitHub_page}). To make the best use of it, a thorough description of the mathematical background and implementation details are given in Section~\ref{S:extraction}. In Section~\ref{S:physics}, we emphasize the importance of the established image processing framework to capture non-conventional field emission behaviour of semiconducting nanodiamond and carbon nanotube materials. Semiconductors and semimetals have long been known to violate the classical field emission law.\cite{1,absorb,fairspace,11,minoux,14,locally} Our approach makes it possible to establish quantitative relation between the applied electric field and experimental field emission current density and therefore allows to verify theory against experiment directly. Results and findings of this work are summarized in Section~\ref{S:Conclusions}.

\section{The image processing algorithm}\label{S:extraction}
\subsection{Feature Extraction}\label{S:FE}

\begin{figure*}
	\centering\includegraphics[scale=0.23]{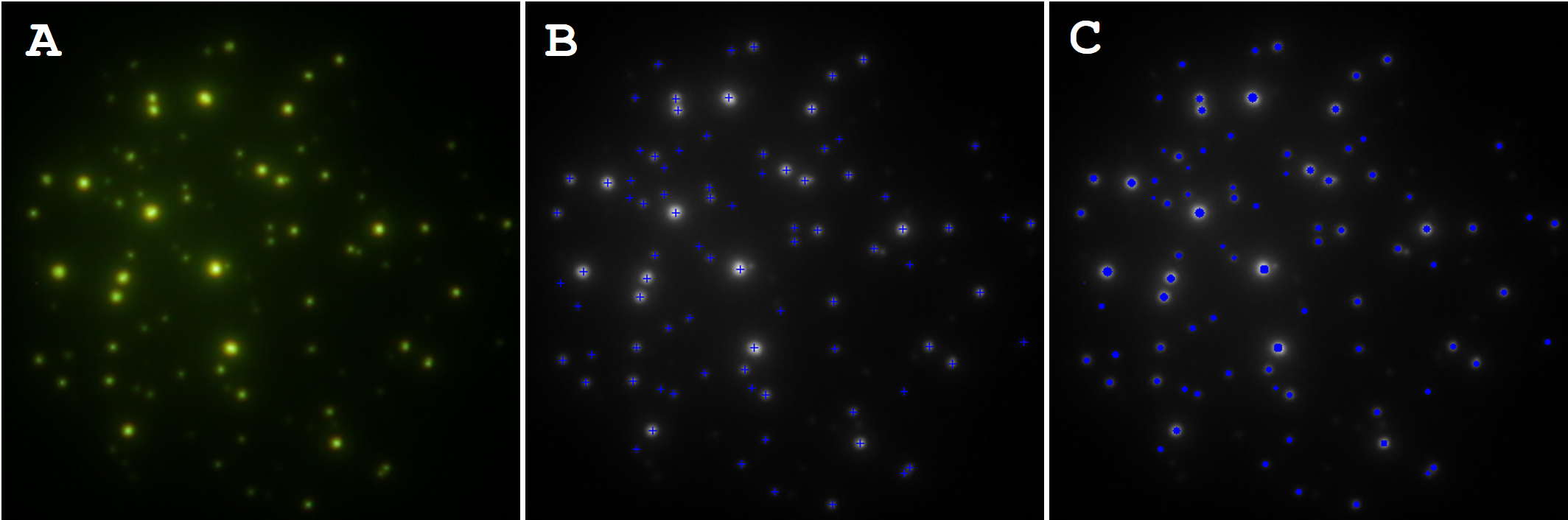}
	\caption{A) A typical micrograph obtained from an UNCD film under the applied dc field.\cite{locally} A 450$\times$450 px$^2$ image represents a projection of spatial distribution of electron emission sites onto a YAG anode placed 106 $\mu$m from a 4.4 mm diameter cathode. B) Detected LMs (shown with blue plus signs) overlaid with emission spots shown on the micrograph. C) Emission pixels (shown in blue), representing the projected emission area, overlaid with emission spots shown on the micrograph.}\label{F:uncdimg}
\end{figure*}

\begin{figure*}
	\centering\includegraphics[scale=0.23]{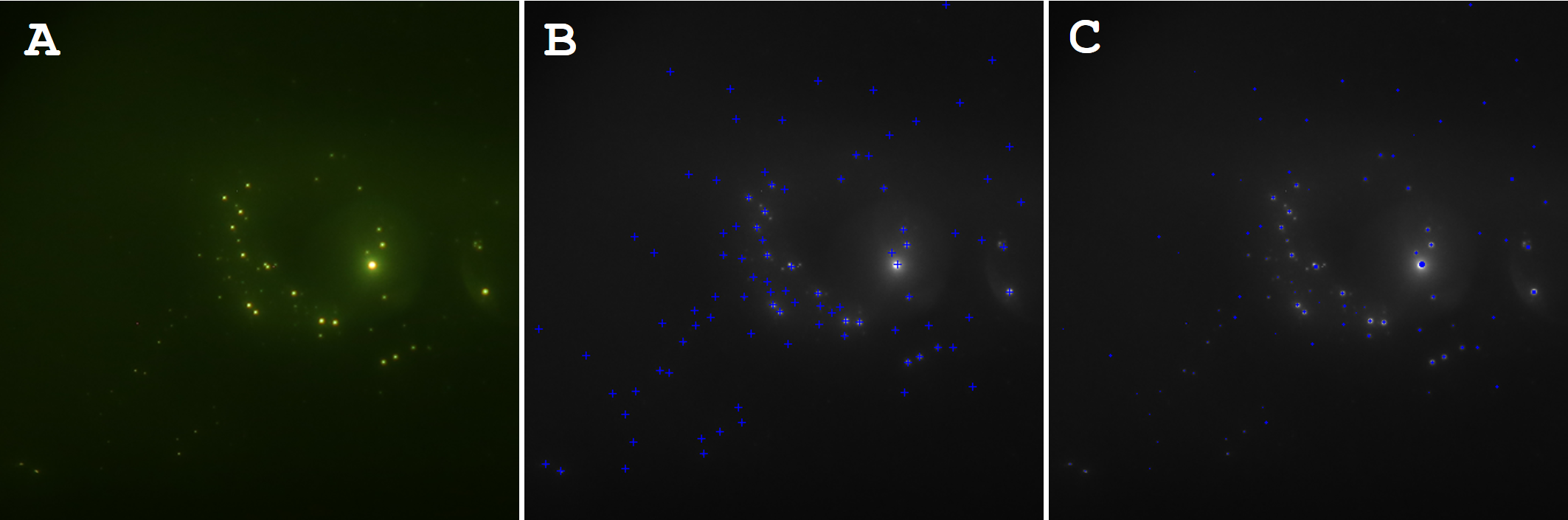}
\caption{A) A typical micrograph obtained from a CNT fiber under dc field. B) Detected LMs (shown with blue plus signs) overlaid with emission spots shown on the micrograph. C) Emission pixels (shown in blue), representing the projected emission area, overlaid with emission spots shown on the micrograph.}\label{F:cntimg}
\end{figure*}

As shown in the examples below (Figs.~\ref{F:uncdimg}A, \ref{F:cntimg}A, and \ref{F:cnt8}A), field emission patterns can vary a lot, with an additional challenge being a bright halo background, which complicates image analysis. However, after a closer look at these images one notices that the patterns consist of bright spots separated spatially enough to distinguish them visually. So, the strategy is to detect the brightest pixel within each emission spot, which is further referred to as a local maximum (LM). Then the number of electron emission sites is equal to the LM count and the emission area can be estimated by assembling certain neighbor pixels around LMs.

\begin{figure}[!b]
	\centering\includegraphics[scale=0.30]{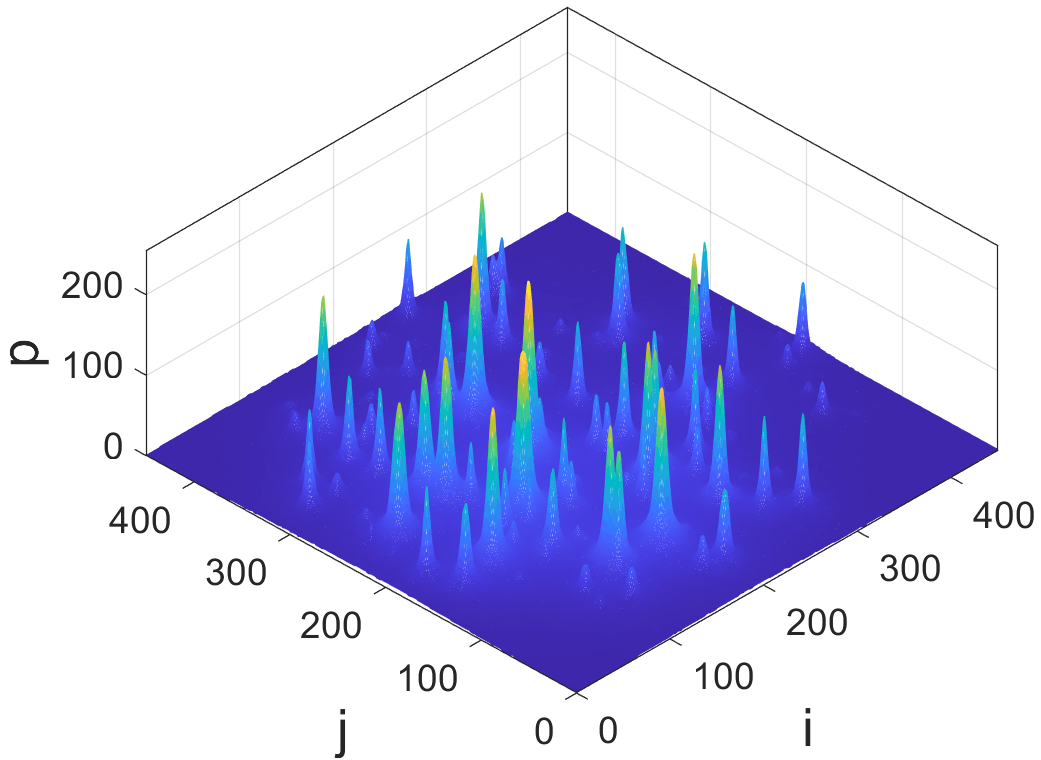}
	\caption{A 3D plot of the micrograph shown in Fig.~\ref{F:uncdimg}A.}\label{F:3dgauss}
\end{figure}

Prior to numerical analysis, an RGB micrograph is converted into a gray-scale format and represented by a 2D matrix of pixels. The intensity of each pixel is given by an integer between 0 and 255. When a typical micrograph (Fig.~\ref{F:uncdimg}A) is represented as a 3D plot (Fig.~\ref{F:3dgauss}), it can be seen that emission spots appear as Gaussian peaks atop a certain background. LMs are the brightest pixels of each Gaussian peak and they are also well separated in space. Therefore, two features must be known for each pixel in order to define LMs: pixel value and the distance to the nearest brighter pixel. Because pixels are represented by integers, occurrence of more than one LM for each peak is possible. In order to prevent this, a small random noise between 0 and 0.1, excluding 0 and 0.1, is added to the image so that there are no identical pixels in the data array. This procedure does not perturb the image because the original image can be retrieved any time by rounding the matrix.

A pixel $ a $ with spatial coordinates $ (i_{a},j_{a})$, where $ i_{a} $ is the row number and $ j_{a} $ is the column number in a 2D array of digitized image, will be represented by $ (p_{a},d_{a}) $ in a feature space, where $ p_{a} $ is the intensity feature and $ d_{a} $ is the distance feature. $p_{a}$ is just a pixel value (an integer, typically between 0 and 255), which can be simply extracted from a data array (before adding the noise).

A fast method of extracting the distance feature is searching for a brighter neighbor in a certain neighborhood of each pixel, which is called the search region\cite{peaksearch, maxfilter}. The search region must be large enough to enclose an entire Gaussian-like peak, but small enough to enclose not more than one entire Gaussian peak. All images presented in this work were 450$\times$450 px$^2$. Although the size of emission spots varied, two standard deviations of each Gauss peak corresponded to 10 pixels or less. At the same time, the distance between peaks was 20 pixels or more. Therefore, a circular search region was chosen: centered around each pixel, the search region radius was set to 10 pixels. It should be noted that for different images or image sizes, this value may have to be adjusted accordingly. The Euclidean distance between any two pixels $a'$ at $(i_{a'}, j_{a'})$ and $a$ at $(i_a, j_a)$ is defined as:

\begin{equation}\label{E:distance}
d_{a'a}=\sqrt{\left( i_{a'}-i_a\right)^{2}+\left( j_{a'}-j_a\right)^{2} }.
\end{equation}
If $s_a$ is the search region for a pixel $ a $ carrying value $p_a$, and a pixel $ b $ carrying value $p_b$ contained inside the $s_a$ is the next closest pixel to the pixel $ a $  such that $p_b>p_a$, then the distance $d_a$ assigned to the pixel $ a $ is
\begin{equation}\label{E:normaldistance}
d_a=\sqrt{\left( i_a-i_b\right)^{2}+\left( j_a-j_b\right)^{2} }.
\end{equation}
\noindent If no other brighter pixel was found in a search region, the brightest pixel is called the maximum in search region (MISR). There is a distance property that can be defined for a MISR as the distance to another closest MISR that is brighter than the former. Say, pixel $A$ with value of $p_A$ is the MISR in its own search region $s_A$, and pixel $B$ with value of $p_B$ is the MISR in its own search region $s_B$, and $p_B > p_A$ while $p_B$ is a MISR closest to pixel $A$, then the distance $d_A$ assigned to pixel $A$ reads
\begin{equation}\label{E:lmdistance}
d_A = \sqrt{\left( i_A-i_B\right)^{2}+\left( j_A-j_B\right)^{2} }.
\end{equation}
The brightest pixel across the entire image is the global maximum (GM). There is a distance value that is assigned to the GM. It is the distance to closest MISR (regardless of its value). Consider, pixel $ GM $ with value of $p_{GM}$ is the global maximum and pixel $ A $ with value of $p_A $ is a MISR, closest to $ GM $, the distance assigned to the $ GM $ is
\begin{equation}\label{E:globaldistance}
d_{GM} = \sqrt{\left( i_{GM}-i_A\right)^{2}+\left( j_{GM}-j_A\right)^{2}}.
\end{equation}

\begin{figure*}
	\centering\includegraphics[scale=0.23]{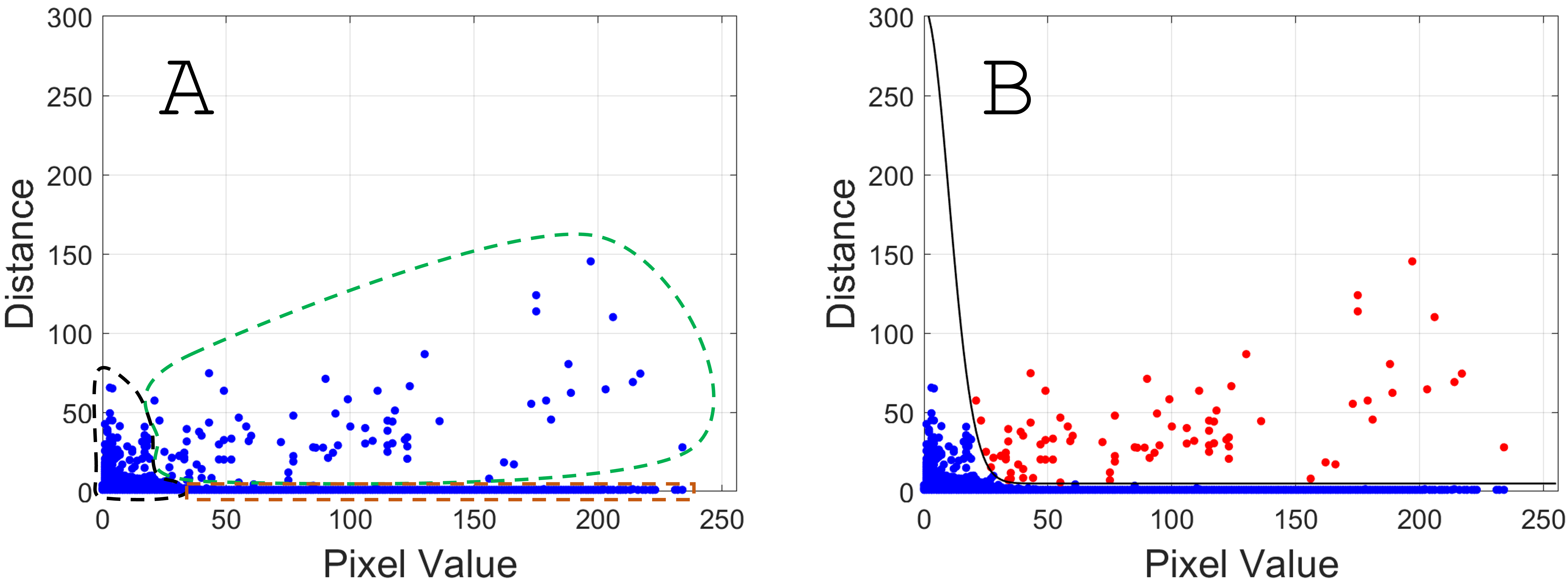}
\caption{A) Decision plot of extracted features for the micrograph obtained from an UNCD cathode operated under the dc field. Green, black, and brown dashed regions show locations of LMs, background pixels, and emission pixels in the decision plot, respectively. There is no overlapping between an LM cluster and a uniform background. B) A black curve shows the Gaussian decision boundary. The pixels shown in red are classified as LMs.}\label{F:uncddb}
\end{figure*}

\begin{figure*}
	\centering\includegraphics[scale=0.23]{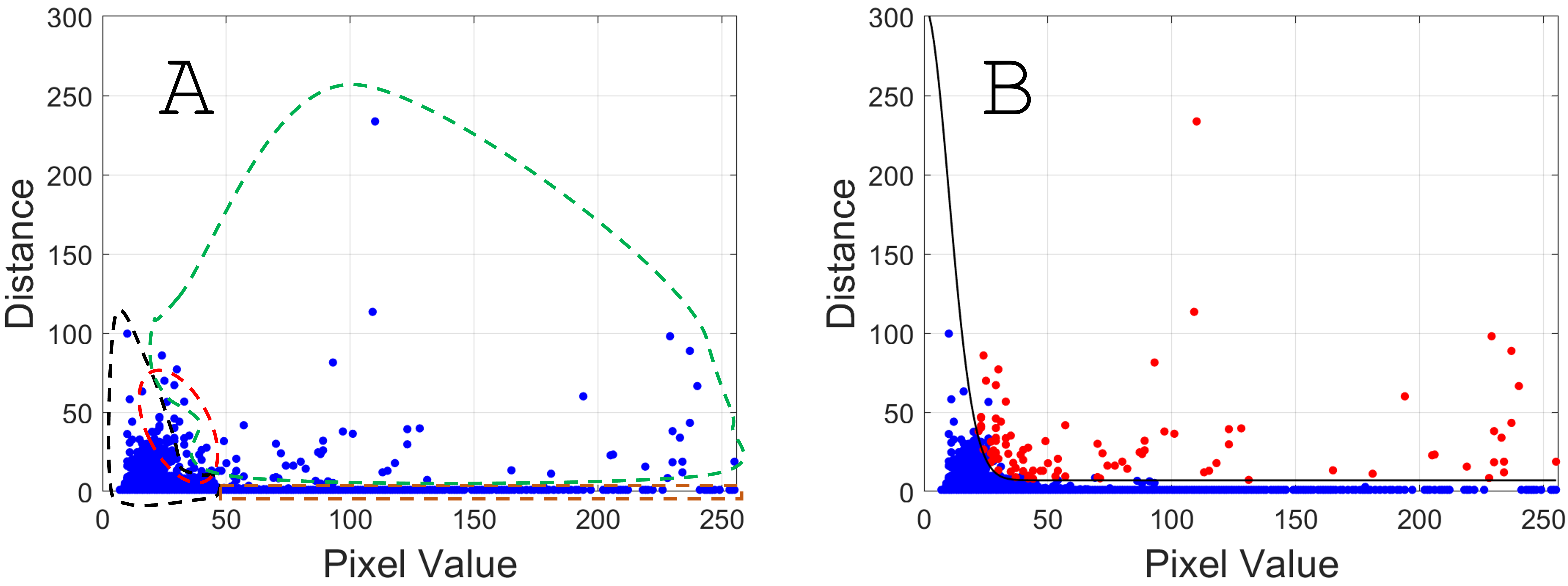}
	\caption{A) Decision plot of extracted features for the micrograph obtained from a CNT fiber under the dc field. Green, black, and brown dashed regions show locations of LMs, background pixels, and emission pixels, respectively. The red dashed region shows overlap between LMs and a background cluster. B) A black curve is the applied Gaussian decision boundary. Red points are detected LMs. Notice that not all points above the boundary are red; this is because some false LMs were filtered out.}\label{F:cntdb}
\end{figure*}

After all these distances are extracted and recorded, the random noise added earlier gets removed. The code stores both $(p_a,d_a)$ and $ (i_{a},j_{a}) $ pairs for every pixel. Scatter plots of intensity--distance arrays, which correspond to emission micrographs from UNCD (Fig.~\ref{F:uncdimg}A) and CNTs (Fig.~\ref{F:cntimg}A), are presented in Figs.~\ref{F:uncddb}A and \ref{F:cntdb}A, respectively. Such plots are called $decision$ $plots$, also known as $plots$ $of$ $features$ in machine learning literature.

\subsection{Decision Boundary}\label{S:DB}

Both the density of emission spots and their intensity increase with applied electric field. Moreover, the background (often non-uniform) arises due to a large number of bright glowing spots. As a result, an analysis of decision plots becomes more complicated. For example, two peaks can be located very close to each other, so the distance feature of a fainter peak becomes very close to or lower than the set distance boundary. Or, let's say, there are fairly faint but distinct emission spots atop a bright background. In this case, there is no clear boundary between the intensity of emission spots and the intensity of background pixels. Moreover, some background protrusions can be mistaken as LMs. Therefore, an effective method for estimating an appropriate decision boundary is required to identify LMs on the decision plot.

A supervised machine learning using labeled data cannot be used for this class of problems because labeling hundreds of images with hundreds of emission spots is not feasible. In addition, micrographs vary a lot for different materials, geometries, or evolve with a high dynamic range throughout a single experiment. An unsupervised machine learning scheme might be used. However, as seen from Figs.~\ref{F:uncddb}A and \ref{F:cntdb}A, no distinct clusters form on the decision plots. Therefore, a simple and robust approach using a tunable decision boundary of a certain form is attempted here.

For pixels that come from a continuous low background (intense and high-gradient background case will be discussed later), the algorithm finds a brighter pixel less than a few pixels away. Thus, these background pixels are characterized by a small distance $d_a$ and a small intensity $p_a$.
As a rule, the background pixels outnumber the emission-spot pixels. Therefore, most of the background pixels lie near origin and appear as crowded clusters encircled by black dashed loops in Figs.~\ref{F:uncddb}A and \ref{F:cntdb}A. Some pixels that belong to emission spots are also characterized by small distances (less than the radius of the search region), but most of them are brighter than the background and their pixel values are large, up to 255. Therefore, these pixels form stretched clusters at the bottom of decision plots as shown by brown dashed regions in Figs.~\ref{F:uncddb}A and \ref{F:cntdb}A.

LMs are bright pixels separated from brighter pixels by large distances (usually larger than the search radius). They are shown by dashed green regions in decision plots. If the background is low and uniform (as in the case shown in Fig.~\ref{F:uncdimg}A), LMs can be easily separated from background pixels (see the green dashed region in a corresponding decision plot in Fig.~\ref{F:uncddb}A). However, when a stronger background appears with a distinct gradient across the image plane as shown in Fig.~\ref{F:cntimg}A, there are relatively faint emission pixels that are barely noticeable on a bright background. In Fig.~\ref{F:cntdb}A, these pixels are shown inside the red dashed region and are intermixed with background pixels. To separate LMs from the background, we use a Gaussian decision boundary given by
\begin{equation}\label{E:gauss}
f(p_a)=A\cdot e^{-\frac{(p_a-\mu)^{2}}{2\sigma^{2}}} + k,
\end{equation}
where $ A $ is the amplitude, $ \mu $ is the mean, $ \sigma $ is the standard deviation, and $ k $ is the offset. Functions, which define decision boundaries for micrographs in Figs.~\ref{F:uncdimg}A and \ref{F:cntimg}A, are shown in Figs.~\ref{F:uncddb}B and \ref{F:cntdb}B, respectively. The decision rule is as follows. Consider a pixel $ a $ carrying the pair of values $ (p_a,d_a) $. If $ d_a>f(p_a) $, then the pixel $ a $ is a LM. The parameters $A$, $\mu$, $\sigma$, and $k$ can be adjusted for each data set. If the background and LM pixels are mixed as shown by the red dashed region in Fig.~\ref{F:cntdb}A, many background pixels are detected as LMs. One way to fix it is to increase $ \sigma $ to exclude background pixels from the LM list. In a case when the density of emission spots is high (emission spots, and thus LMs, are spatially very close to each other), decreasing $ k $ helps to identify some missing LMs. The distance feature $d_a$ was never greater than 300 for 450$\times$450 px$^2$ images. Thus, the parameter $A$ was kept at the value of 300 for all data sets presented. Mean $ \mu $ was set to zero for all data sets. Any false LM can be further filtered out by applying the surface-fitting method given in Sec.~\ref{S:area}. Therefore, a crude adjustment of the decision boundary parameters is enough.

Decision boundaries shown with black curves in Figs.~\ref{F:uncddb}B and \ref{F:cntdb}B were applied to Figs.~\ref{F:uncddb}A and \ref{F:cntdb}A, respectively. Then, false LMs were filtered out by the surface-fitting method. A final list of detected LMs are labeled with red data points on the decisions plots in Figs.~\ref{F:uncddb}B and \ref{F:cntdb}B as well as with blue crosses in Figs.~\ref{F:uncdimg}B and \ref{F:cntimg}B. These examples illustrate nearly perfect LM detection.

\subsection{Emission Area}\label{S:area}

As it was already mentioned, each emission spot on a micrograph (e.g., Fig.~\ref{F:uncdimg}A) appears on a 3D plot (Fig.~\ref{F:3dgauss}) as a quasi-symmetric Gaussian with its center at a LM. Once a LM is detected, a 2D Gaussian function can be used to fit an intensity peak\cite{mpfit, blobdetection}. The fitting function, which returns the estimated pixel value $ p_{ij}^{e} $ for a pixel at the position $ (i,j) $, is given by 
\begin{equation}\label{E:gaussfit}
p_{ij}^{e}=A\cdot e^{-\frac{(i-i_{LM})^2+(j-j_{LM})^2}{2\sigma^2}}+C,
\end{equation}
where $ A $ is the amplitude, $ \sigma $ is the standard deviation, $C$ is the offset to manage the background level, and $(i_{LM},j_{LM})$ is the position of a given LM. Fitting parameters $ A $, $ \sigma $, and $ C $ have to be determined for each LM. A 15$\times$15 px$^2$ region was chosen to fit each peak centered at the position $(i_{LM},j_{LM})$, so $ i_{LM}-7 \leq i \leq i_{LM}+7 $ and $ j_{LM}-7 \leq j \leq j_{LM}+7 $. Such fitting region is optimal for our images  because it ensures that only one emission spot is included. The fitting parameters for each LM can be calculated via minimizing the following sum by the least squares regression method
\begin{equation}\label{E:leastsq}
\sum_{i}\sum_{j} (p_{ij}-p_{ij}^e)^2,
\end{equation}
where $ p_{ij} $ is the original pixel value, $ i $ and $ j $ are the indices running over the fit region of each LM.

Obviously, most of the bright pixels will fall within one standard deviation of the mean. To classify a pixel $a$, which is located at $(i_a,j_a)$ within a fit region centered at a LM at $(i_{LM},j_{LM})$, as the $emission$ $pixel$ that contributes toward the emission area calculation, the following condition must be satisfied

\begin{equation}\label{E:emssionpixel}
(i_a-i_{LM})^2+(j_a-j_{LM})^2<\sigma^2.
\end{equation}
Alone, this condition is not enough to classify a pixel as the emission pixel. For example, consider the case when a background pixel is incorrectly detected and listed as a LM (we call it a false LM in Sec.~\ref{S:DB}). Gaussian fit of the region around such a false LM will yield a large standard deviation $\sigma$, but small amplitude $A$. This allows for filtering out fake LMs by applying threshold values for $A$ and $\sigma$. LMs, which satisfy the condition
\begin{equation}\label{E:fitcondition}
(\sigma>\sigma_{th})\lor(A<A_{th}),
\end{equation}
are not real LMs and they are discarded from the master LM list. Thus, they do not contribute to the local maxima count and the emission area calculation.

For the micrographs presented in this work, $\sigma_{th}=7$ and $A_{th}=10$ are chosen. For cases when spots are faint, $A_{th}$ should be decreased to make the spots detected. If background noise is high, $\sigma_{th}$ should be decreased to produce reliable results. In Figs.~\ref{F:uncdimg}C and \ref{F:cntimg}C, the blue pixels identified as the emission pixels are overlaid with the emission spots shown in Figs.~\ref{F:uncdimg}A and \ref{F:cntimg}A, respectively. If the physical area of the image is known (we typically work with square images with a side length equal to the diameter of the cathode), then the apparent emission area is given by

\begin{equation}\label{E:earea}
\begin{split}
\text{Emission area (mm$^2$)} &= \text{Image area (mm$^2$)}\\
&\quad\times\frac{\text{Number of emission pixels}}{\text{Total number of pixels}}.
\end{split}
\end{equation}

Both Figs.~\ref{F:uncdimg}A and \ref{F:cntimg}A have 450$\times$450=202,500 pixels. There are 1,929 emission pixels in Fig.~\ref{F:uncdimg}C and 349 emission pixels in Fig.~\ref{F:cntimg}C. The displayed portion of a YAG:Ce screen in Figs.~\ref{F:uncdimg}A and \ref{F:cntimg}A has the area of 4.4$\times$4.4 mm$^2$ and 12.4$\times$12.4 mm$^2$, respectively. Then applying Eq.~\ref{E:earea}, the apparent emission area is 0.814 mm$^2$ for UNCD (Fig.~\ref{F:uncdimg}) and 0.264 mm$^2$ for CNTs (Fig.~\ref{F:cntimg}), while the total cathode surface area available for emission is 15.2 mm$^2$ for UNCD and 4.71 mm$^2$ for CNTs.

\subsection{Micrographs with Intense Background }\label{S:intense}

\begin{figure*}
	\centering\includegraphics[scale=0.23]{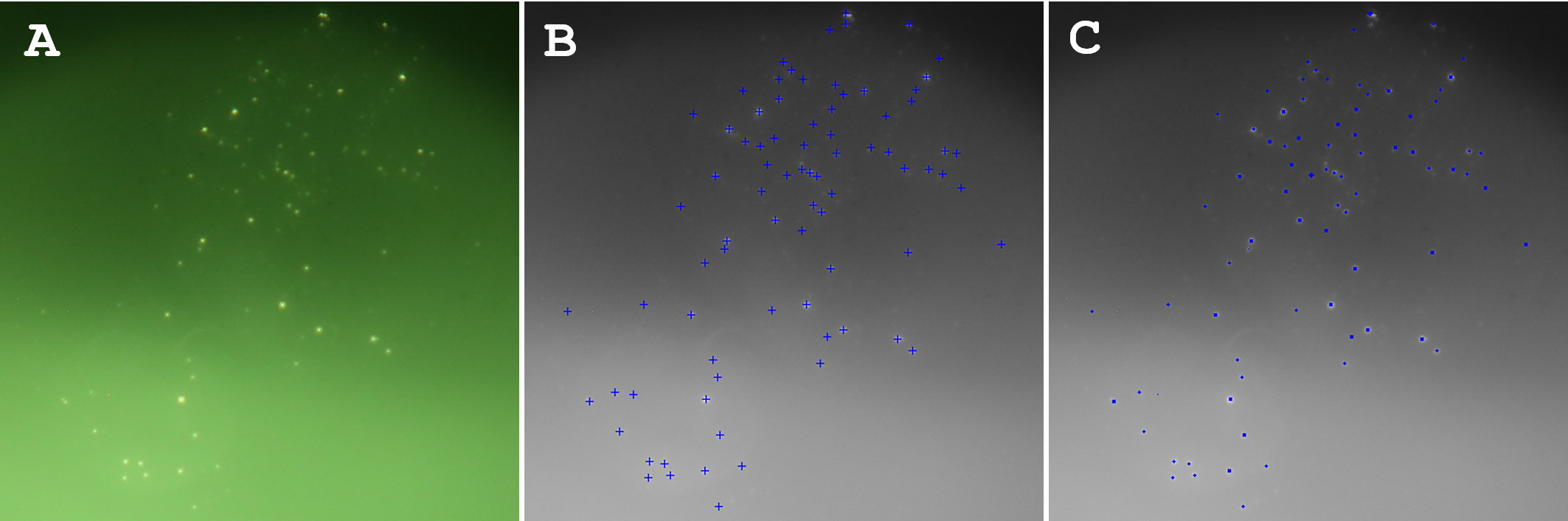}
		\caption{A) A typical micrograph obtained from a CNT fiber under dc field with glowing high-gradient background. The emission domains appear as bright peaks on the glow. Exact source of the glow is unknown. B) Detected LMs (shown with blue plus signs) overlaid with emission spots shown on the micrograph. C) Emission pixels (shown in blue), representing the projected emission area, overlaid with emission spots shown on the micrograph.}\label{F:cnt8}
\end{figure*}

\begin{figure*}
	\centering\includegraphics[scale=0.23]{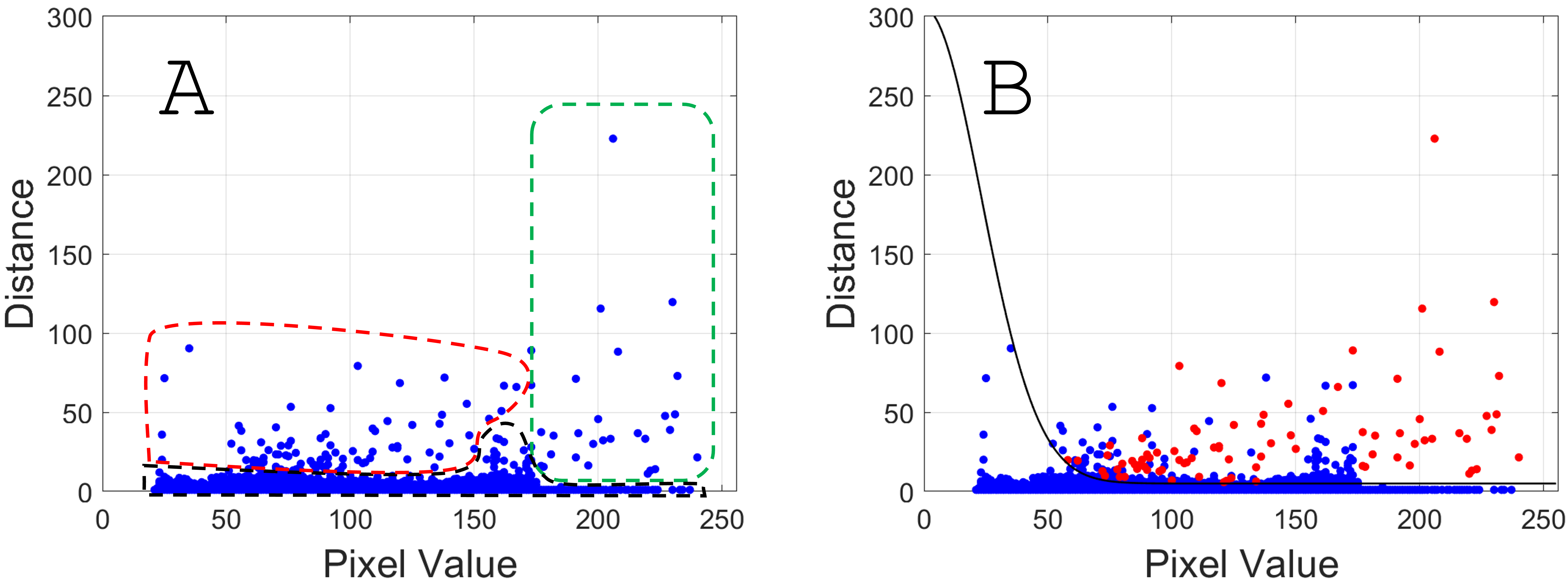}
\caption{A) Decision plot of extracted features for the micrograph obtained from a CNT fiber under the dc field with glowing high-gradient background. Unlike the previous cases, the background does not form a cluster. Instead, it is distributed over a wide pixel range shown in a black dashed region. Green dashed region consists of well separated LMs. Red region includes both background pixels and LMs, so the boundary should be drawn so that that region includes candidate LMs. Any false LMs are to be filtered out by Gaussian surface fitting. B) A black curve shows the applied decision boundary. All points above the curve are candidate LMs. Red points label finalized LM list after false LMs are filtered out by the surface fitting.}\label{F:cnt8db}
\end{figure*}

Micrographs with intense and non-uniform backgrounds are extremely challenging to process and analyze. One such example is shown in Fig.~\ref{F:cnt8}A. In this micrograph, the global background changes from dark at the top to very bright at the bottom, where the detection of emission spots becomes less effective. A corresponding decision plot is shown in Fig.~\ref{F:cnt8db}A. Unlike previous cases (Figs.~\ref{F:uncddb} and \ref{F:cntdb}), the background pixels do not cluster near the origin, but instead they are distributed over a wide range of pixel values (shown by a black dashed region in Fig.~\ref{F:cnt8db}A). There are well separated bright pixels, corresponding to LMs, shown by a green dashed region. Because some of the glowing background pixels have values comparable or higher than those of some of the emission pixels, there is also a large region, shown by a red dashed line, where LMs and background pixels are intermixed.

It is important to note that any LM, excluded during the decision-making procedure, cannot be recovered at later stages of the image analysis. However, any background pixel, identified and included as a LM, can be filtered out by the surface-fitting procedure. Therefore, the decision boundary should be $soft$, i.e. it should include all possible LM candidates (as opposed to a $rigid$ boundary that would exclude as many background pixels as possible). According to this approach, the decision boundary should be drawn on the decision plot (Fig.~\ref{F:cnt8db}) in such a way that the pixels from a red region are considered as LM candidates. Any false LM will be likely filtered out. One major drawback of applying the soft boundary is the increased computation time due to increased number of Gaussian fitting steps required for increased number of LMs. With the decision boundary applied in Fig.~\ref{F:cnt8db}B, all pixels above the black curve are LM candidates. Fig.~\ref{F:cnt8db}B highlights in red a finalized array of LMs (after most or all false LMs were removed). Same points are overlaid on the original image as blue crosses in Fig.~\ref{F:cnt8}B. Nearly perfect agreement can be seen upon visual inspection. Finally, Fig.~\ref{F:cnt8}C shows in blue the calculated apparent emission area. One can see how effectively the entire pattern recognition workflow performs even for images highly distorted by the background. The calculated number of emission pixels is 533 out of 202,500 pixels. The field of view in Fig.~\ref{F:cnt8}A is 11.2$\times$11.2 mm$^2$ yielding an emission area of 0.33 mm$^2$ while the total cathode area is 15.2 mm$^2$.

\section{Results and Discussions}\label{S:physics}

\begin{figure*}
	\centering\includegraphics[scale=0.4]{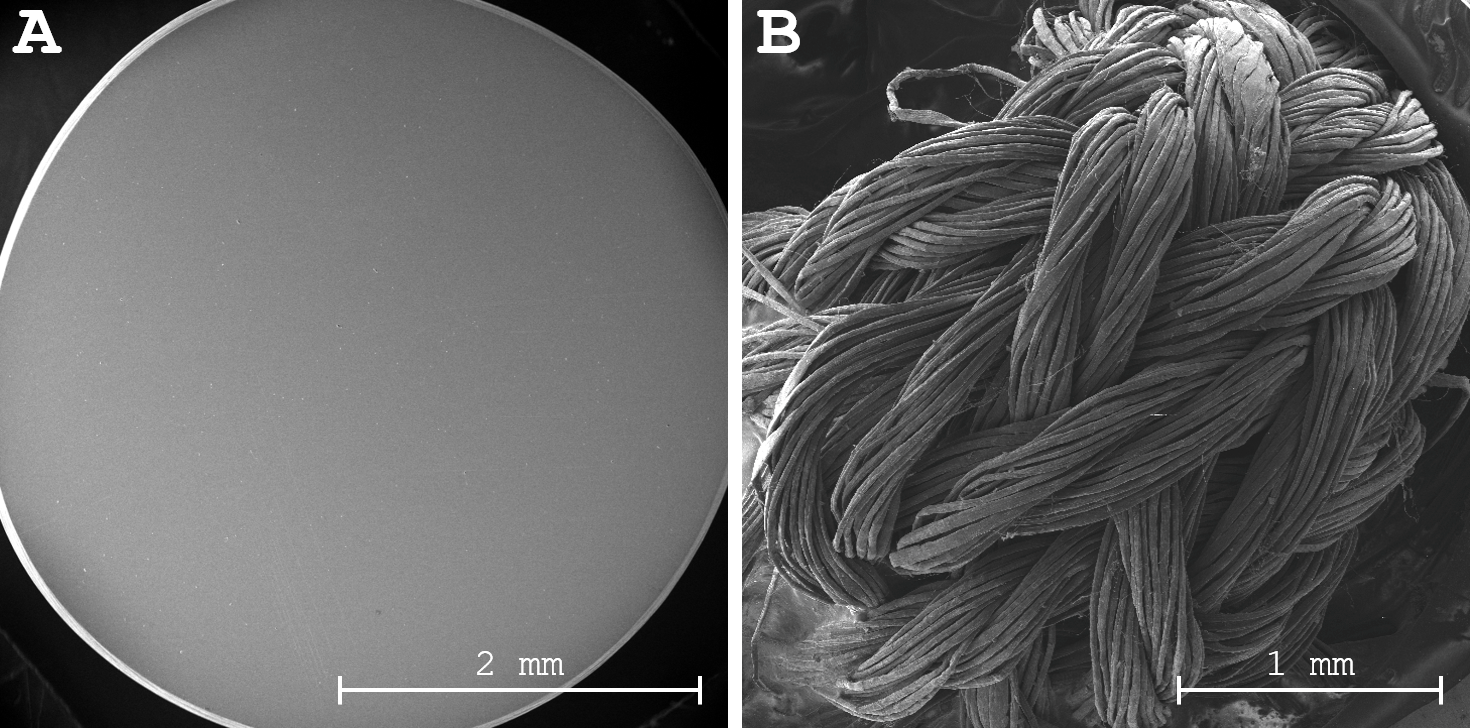}
	\caption{A) SEM image of the UNCD film grown on a Ni/Mo/SS substrate. B) SEM image of the CNT fiber. The twisted and folded yarn structure is clear from the image.}\label{F:samples}
\end{figure*}

In this section, we use the described algorithm to extract the electron emission area projected on the YAG screen, or the apparent emission area, and estimate the experimental current density as a function of the electric field. We analyze three sets of field emission micrographs. Each set contains tens of micrographs taken with regular intervals as the applied voltage (and the electric field) was swept up and down. The current was measured concurrently. The first set is obtained from the $n$-type UNCD film grown on a planar stainless steel (SS) substrate with a Ni/Mo buffer layer\cite{locally}. The second and third sets are from the CNT folded rope taken before and after conditioning, respectively. This sample was formed by braiding and twisting 90~$\mu$m fibers, then folding the resulting yarn multiple times and inserting it in a metal tube.\cite{mypaper} Scanning electron microscopy (SEM) images of the samples are given in Fig.~\ref{F:samples}.

\begin{figure*}
	\centering\includegraphics[scale=0.4]{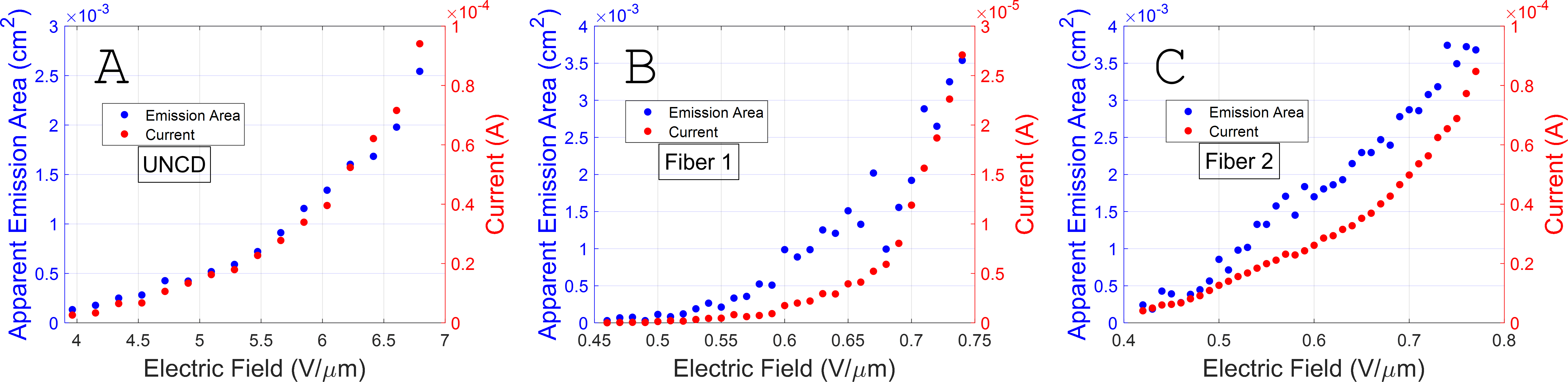}
	\caption{The apparent emission area estimated using the algorithm (blue dots) and the measured current (red dots) as a function of the applied electric field A) for the UNCD film, B) for the CNT fiber before conditioning, and C) for the CNT fiber after conditioning.}\label{F:earea}
\end{figure*}

Electric field dependencies of the apparent emission area $S_{\text{apparent}}$ computed  for the considered data sets are presented with blue dots in Fig.~\ref{F:earea}. In the same figure, red dots trace corresponding field dependencies of the measured current $I_{\text{measured}}$. As it was already mentioned above, the apparent emission area is not the actual emission area. The apparent emission area estimated by the algorithm is the projected, or magnified, emission area. The magnification occurs as a result of the non-zero finite transverse momentum of emitted electrons as well as the transverse component of the fringe electric field on the surface of the high-aspect-ratio features. Therefore, the relation between the apparent emission area and the real emission area can be obtained by introducing the magnification factor. Although finding the factor is not a straightforward process (because it depends on both internal material parameters and complicated surface geometry), the lower and upper limits of the experimental current density can be easily estimated.

\begin{figure*}
	\centering\includegraphics[scale=0.43]{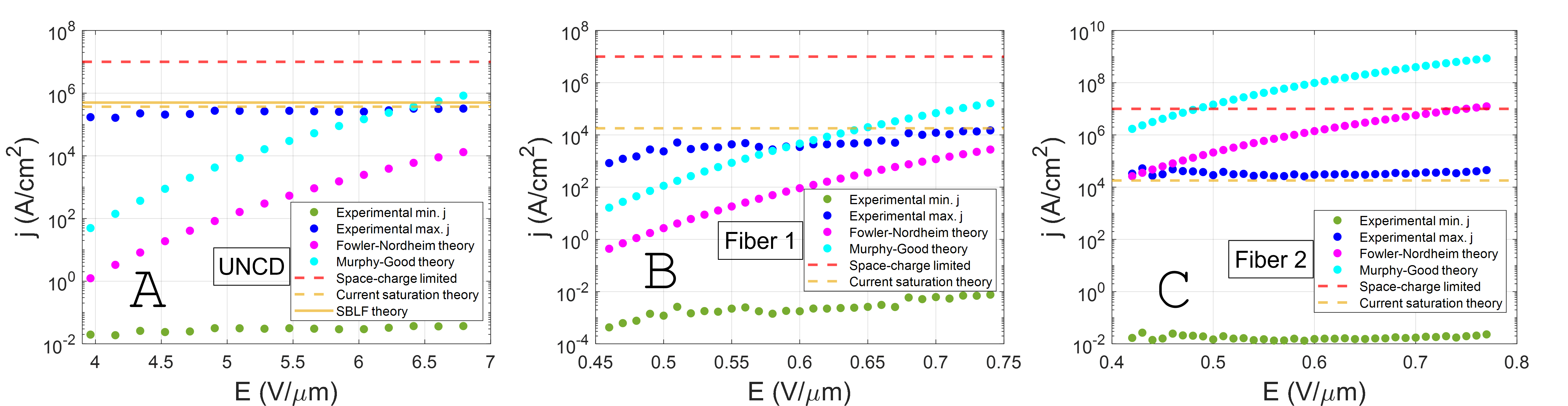}
	\caption{The experimental current densities (the lower limit is shown with green dots; the upper limit is shown with blue dots) compared with the theoretical estimations (the elementary FN equation is shown with magenta dots; the Murphy-Good extension is shown with cyan dots) as well as the space-charge limited emission (dashed red line) and the saturation-velocity limited emission (dashed orange line) as a function of the applied electric field A) for the UNCD film, B) for the CNT fiber before conditioning, and C) for the CNT fiber after conditioning.}\label{F:physics}
\end{figure*}

The lower limit of the current density can be found by dividing the measured current $I_{\text{measured}}$ by the total apparent emission area $S_{\text{apparent}}$ estimated by the algorithm for a given electric field $E$
\begin{equation}\label{jmin}
    j_{\text{exp.min}} = \frac{I_{\text{measured}}}{S_{\text{apparent}}}.
\end{equation}
The lower current-density limits $j_{\text{exp.min}}$ for all three data sets are shown with green dots in Fig.~\ref{F:physics}. It can be seen that the current density $j_{\text{exp.min}}$ does not change significantly with the applied electric field. This happens because the variation of the apparent emission area with the applied electric field is almost identical to that of the measured current (Fig.~\ref{F:earea}). Such behaviour is observed for all data sets and contradicts to the Fowler-Nordheim law,\cite{fntriangle} which predicts the exponential increase of the current density with the applied electric field.

Further on, the upper limit of the current density can be calculated using an estimated angle of the beam ray trajectory. If the angle is $\alpha$, then a single emission site/domain with area  $S_\text{single}$ appears on the YAG:Ce screen as $S_{\substack{\text{single}\\ \text{apparent}}}$ given by
\begin{equation}\label{}
    S_{\substack{\text{single}\\ \text{apparent}}}=\pi(d\tan{\alpha})^2,
\end{equation}
where $d$ is the inter-electrode gap. Geometrically, the angle $\alpha$ can be obtained using the mean transverse energy (MTE) as\cite{mypaper}
\begin{equation}\label{}
    \tan{\alpha}=\frac{p_x}{p_z}=\sqrt{\frac{2\cdot\text{MTE}}{m_0c^2}}\frac{1}{v/c}\frac{1}{\gamma},
\end{equation} 
where $m_0c^2= 0.511\ \text{MeV}$ is the electron rest energy, $v/c$ is the ratio of electron speed to speed of light, and $\gamma$ is the Lorentz factor. At $1$~keV energy, $v/c$ and $\gamma$ are taken as $0.063$ and $1$, respectively. MTE values are 100-200 meV\cite{MTE_UNCD_1, MTE_UNCD_2} for the UNCD; and for CNT its Fermi energy of 50-100 meV\cite{cnt_energy} can be used. Then, the estimated values of the angle is $0.9^{\circ}$ for the UNCD sample\cite{oksana_theory} and $0.4^{\circ}$ for the fiber sample\cite{mypaper}. Then the inverse magnification factor is given by the ratio $S_{\text{single}}/S_{\substack{\text{single}\\ \text{apparent}}}$ and the current density can be calculated as
\begin{equation}\label{E:upperlimit}
\begin{split}
    j_{\text{exp.max}}=\frac{I_{\text{measured}}}{\frac{S_{\text{single}}}{S_{\substack{\text{single}\\ \text{apparent}}}}\cdot S_{\text{apparent}}}.
\end{split}
\end{equation}

To obtain the upper current density limit, the minimum possible size for $S_{\text{single}}$ should be chosen in Eq.~\ref{E:upperlimit}. Therefore, we assume $S_{\text{single}}$ to be equal to the size of grain boundaries ($1$~nm by $1$~nm square\cite{oksana_theory}) and the size of a single CNT ($10$~nm in diameter circular region) for the UNCD and the fiber, respectively. The resulting upper limits for the three datasets are shown with blue dots in Fig.~\ref{F:physics}. A few important conclusions can be made. First, experimental current densities are practically constant in the studied range of electric fields. Therefore, an increase in current with electric field can be explained by an increase in emission area (see Fig.~\ref{F:earea}). Second, according to Dyke $et$~$al.$\cite{spacecharge}, above $10^7 \text{A}/\text{cm}^2$ current density level, the emission is limited by the space charge, which causes decrease in effective field over the cathode. The current density limit due to the space charge is shown by a red dashed line for reference in Fig.~\ref{F:physics}. However, we find that the upper limit of experimental current density is much lower than this value. Therefore, the saturation behaviour cannot be explained by the space-charge effect.

We also compare our results with predictions given by the field emission theory in both Fowler-Nordheim (FN) and Murphy-Good (MG) formulations. In FN formulation, the image-charge effect is ignored and the potential barrier at the material-vacuum interface can be approximated by a simple triangular potential. In this case, the FN current density $j_{\text{FN}}~(\frac{\text{A}}{\text{m}^2})$ is given by\cite{fntriangle,sommerfeld}
\begin{equation}\label{E:fn}
    j_{\text{FN}}=A \frac{1}{\phi}
    (\beta E)^{2}
    e^{-\frac{B \phi^{3/2}}{\beta E}},
\end{equation}
where $ A=1.54\times 10^{-6}\frac{\text{AeV}}{\text{V}^{2}} $, $ B=6.83\times 10^{9}\frac{\text{V}}{\text{eV}^{3/2}\text{m}} $, $ \beta $ is the unitless field enhancement factor, $ \phi~(\text{eV})$ is the material work function which can be taken as 4.8~eV for both UNCD and fiber samples, and $E~(\frac{\text{V}}{\text{m}})$ is the electric field. $\beta$ can be obtained from experimental data by plotting it in FN coordinates where $x$-axis is $1/E$ and $y$-axis is $\ln(j/E^{2})$. The plot forms a distinct knee point for non-metal and semi-metal emitters\cite{locally,mypaper} that highlights the saturation behaviour in a high-field region. After fitting a low-field region with a line as in Ref.~\onlinecite{mypaper}, the slope $m$ is given by
\begin{equation}\label{}
    m = -\frac{B\phi^{3/2}}{\beta},
\end{equation}
where $\beta$ can be extracted from. The resulting current density curves are shown by magenta dots in Fig.~\ref{F:physics}.

A better theoretical approach is taking the image-force/charge effect into account. In this case, the surface potential barrier is given by a rounded triangular barrier and the current density is given by the MG current density $j_{\text{MG}}~(\frac{\text{A}}{\text{m}^2})$\cite{murphy,houston1,sommerfeld,forbes1,fnimageforce,fursey2005}
\begin{equation}\label{E:mg}
    j_{\text{MG}}=A \frac{1}{\text{t}_\text{F}^2\cdot\phi}
    (\beta E)^{2}
    e^{-\text{v}_{\text{F}}\frac{B \phi^{3/2}}{\beta E}},
\end{equation}
where $A$ and $B$ are the same constants as before, and $\text{t}_{\text{F}}$ and $\text{v}_{\text{F}}$ are special field emission elliptic functions of unitless Nordheim parameter $y$ given by\cite{fnimageforce,houston1}
\begin{equation}\label{E:npar}
  y = k\cdot\frac{1}{\phi}\sqrt{\beta E},
\end{equation}
where $k=3.79\times10^{-5}~\frac{\text{eV}\cdot\text{m}^{1/2}}{\text{V}^\text{1/2}}$, $E$ and $\phi$ are in terms of $\text{V}/\text{m}$ and $\text{eV}$, respectively. Once it is plotted in FN coordinates, it produces the slope $m$ given by
\begin{equation}\label{E:slopemg}
    m = -\text{s}_{\text{F}}\frac{B\phi^{3/2}}{\beta},
\end{equation}
where $\text{s}_{\text{F}}$ is the slope correction factor. Although $\text{v}_{\text{F}}$, $\text{t}_{\text{F}}$, and $\text{s}_{\text{F}}$ are not trivial to calculate, they had been tabulated well\cite{houston2} and simple good approximations are given by\cite{eliptic_approx}
\begin{equation}\label{E:vf}
    \text{v}_{\text{F}}=1-y^2+\frac{1}{3}y^2\ln{y},
\end{equation}
\begin{equation}\label{E:tf}
    \text{t}_{\text{F}}=1+\frac{1}{9}(y^2-y^2\ln{y}),
\end{equation}
\begin{equation}\label{E:sf}
    \text{s}_{\text{F}}=1-\frac{1}{6}y^2.
\end{equation}
For a typical range of applied fields and the work function of 4.8~eV, $\text{s}_\text{F}$ and $\text{t}_\text{F}$ can be adequately taken as $0.95$ and $1$, respectively\cite{houston2,forbessfvf}. Then, $\beta$ can be extracted from Eq.~\ref{E:slopemg} after plotting experimental data in FN coordinates and applying previous procedure. By substituting the extracted $\beta$ in Eq.~\ref{E:npar} and using approximation of the Nordheim function given by Eq.~\ref{E:vf}, $j_{\text{MG}}$ in Eq.~\ref{E:mg} can be calculated. The corresponding theoretical current-density curves are shown by cyan dots in Fig.~\ref{F:physics}. It is obvious that both elementary FN equation (Eq.~\ref{E:fn}) and MG equation (Eq.~\ref{E:mg}) fail to explain the saturation trend observed experimentally.

The discrepancy between theory and experiment is due to the fact that FN type equations are derived for the metal where the number of carriers is approaching infinity, surface is equipotential, and resistance on the current path can be ignored. However, this assumption is obviously not valid for non-metal and semi-metal emitters: compare the carrier concentration $\sim 10^{23}$ cm$^{-3}$ in copper to the carrier concentration $\sim 10^{18}$-$10^{19}$ cm$^{-3}$ in CNT and UNCD. The development of improved theoretical models that would take semiconductor effects into account is critical. Nevertheless, most studies relate this saturation effect to impurities at emitter apex\cite{absorb}, space-charge effect\cite{fairspace,verboncoeur}, series contact resistance between substrate and the emitter\cite{seriesresistance,faircontact}, etc. As we stated before, the space charge is not a dominant effect at the current-density levels of our interest. Surface adsorbates do not cause saturation for metal emitters, so we believe it is not an important effect for our samples. Measurements\cite{minoux} show that voltage drop due to series contact resistance is small for typical range of applied potential, so it can also be ruled out. On the other hand, it is proven both experimentally\cite{minoux} and analytically\cite{forbes2} that small voltage drop along the emitter body can cause large drop in enhancement factor at the emitter apex. The voltage loss along the body can also be modeled by penetration of the field into the sample through defining a certain depletion length of $W$ from the emitter surface\cite{currentsaturation} into the bulk. Such a $W$ is a result of poor screening due to limited number of charge carriers in non-metallic field emitters. Therefore, the effects of the decrease in field enhancement factor as a result of voltage loss along the emitter\cite{forbes2,minoux} or limited charge supply\cite{currentsaturation} appear to be better hypotheses. The saturation current density, i.e. the maximum current density that can be extracted from a sample assuming carriers reach saturation velocity, is given by\cite{currentsaturation}
\begin{equation}\label{E:sat}
    j_{\text{sat}} = \frac{e n^{2/3} v_{\text{sat}}}{W} = \frac{e n^{2/3}}{\tau},
\end{equation}
where $j_\text{sat}~(\frac{\text{A}}{\text{cm}^2})$ is the saturation current density, $e=1.6\times10^{-19}~\text{C}$ is the electron charge, $n~(\text{cm}^{-3})$ is the electron carrier density, $v_\text{sat}~(\frac{\text{cm}}{\text{s}})$ is the saturation velocity, $W~(\text{cm})$ is the length of the sub-surface depletion region, and $\tau$ is the transit time, i.e. the time it takes for charges to cross over the depletion region and reach the cathode surface to get emitted into vacuum. Typical parameters for CNT fiber\cite{mypaper} are: $n\sim 10^{18}~\text{cm}^{-3}$, $v_\text{sat}\sim 10^7~\text{cm}/\text{s}$, and $W\sim 8.9\times 10^{-5}~\text{cm}~(890~\text{nm})$. Then, Eq.~\ref{E:sat} gives $1.8\times10^4~\frac{\text{A}}{\text{cm}^2}$ which is shown with orange dashed line in Figs.~\ref{F:physics}B and \ref{F:physics}C for reference. Another dashed orange line $3.7\times10^5~\frac{\text{A}}{\text{cm}^2}$ is also drawn for UNCD in Fig.~\ref{F:physics}A, where $n$ is taken as $10^{19}~\text{cm}^{-3}$ and $W$ is replaced with the film thickness of 200 nm atop of a metal substrate that would terminate electric field lines. Yet another, orange solid line, is presented in the same plot. This additional line is based on the extended Stratton-Baskin-Lvov-Fursey (SBLF) theory formulated in our previous work (see Ref.~\onlinecite{oksana_theory}). In SBLF theory, Poisson's and Stratton's equations are solved self-consistently at the material-vacuum interface. The extended SBLF takes into account detailed density of states of a material and, importantly, accounts for the effect of field-induced reduction of electron mobility (referred in this work as the transit time limitation) that plays a central role. The comparison between solid and dashed orange lines suggests that Eq.~\ref{E:sat} can be safely used in place of detailed and time consuming SBLF calculations if a fast benchmarking of experimental data is needed.

Fig.~\ref{F:physics} demonstrates the behaviour of real field emitters that is significantly different from that dictated by the classical field emission theory. First of all, we find that neither Fowler-Nordheim nor Murphy-Good formulation of $j$-$E$ relation captures the functional behaviour of the studied field emitters, while demonstrating significant disagreement with experimental measurements. Thus, the classical field emission appears to have no relevance toward describing experimental behaviour of this class of cathodes (be it a first-time turn-on or a consecutive conditioned operation). Additionally, the 1D vacuum space-charge effect (it is obtained from a 1D Child-Langmuir law treatment\cite{spacecharge} and thus provides the lowest boundary for the space-charge saturation limit as compared to a 2D case\cite{2DCL}) still overestimates the experimentally observed saturation by 1 to 3 orders of magnitude.

The transit-time limited field emission model appears to capture the experimentally observed emission characteristics unique to the advanced semiconducting field emission materials, UNCD and CNT. It also correctly explains two major facts important for cathode conditioning ('Fiber1' and 'Fiber 2' case studies in this work). First, despite the characteristic field emission parameters (e.g., field enhancement factor and turn-on field) may drastically alter, the current density saturates due to intrinsic material properties and thus remains the same as long as the material itself does not alter during conditioning. Second, it correctly implies that once a semiconductor is depleted, the increase of the output current is only possible if the emission area increases (see the results in Fig.~\ref{F:earea}). The emission area grows with the electric field and so does the output current, but the current density is a constant of the electric field. This creates a sizeable misconception when measured $I$-$V$ or $I$-$E$ curves, and not $j$-$E$ curves, are analyzed using the elementary Fowler-Nordheim approach.

\section{Conclusions}\label{S:Conclusions}
In conclusion, we demonstrated that the experimental field emission current density of UNCD and CNT materials remained constant with applied electric field demonstrating saturation behaviour. These results emphasized that classical Fowler-Nordheim and Murphy-Good theories fail to describe fundamental field emission properties of these materials. Moreover, it was shown that the saturation level cannot be explained by such external effects like space charge, surface adsorbates, or contact resistance. Instead, it was found that the transit-time limited field emission, which accounts for the limited carrier concentration and the saturation drift velocity, provides the current-density saturation levels very close to the upper limits of experimentally measured ones for all cases studied in this work.

The direct benchmarking between theory and experiment became possible solely thanks to the development of a new pattern recognition algorithm. Application of this algorithm was demonstrated and emphasized by counting the number of field emission sites and calculating the apparent field emission area from dc micrographs. The algorithm is fast. Run on a personal laptop with Intel i5 2.5 GHz core, the full set of calculations took 23.8 seconds for Fig.~\ref{F:uncdimg}, 18.9 seconds for Fig.~\ref{F:cntimg}, and 26.3 seconds for Fig.~\ref{F:cnt8}. It took the longest processing time for Fig.~\ref{F:cnt8} because a soft decision boundary was used and false LMs had to be sorted out.

Future work is being focused on unsupervised machine learning, especially one-class support vector machines, to replace the Gaussian decision boundary and thus completely automate the pattern recognition. The current/future releases are/will be available as a freeware on GitHub\cite{GitHub_page}.

\section*{Acknowledgments}
This material is based upon work supported by the U.S. Department of Energy, Office of Science, Office of High Energy Physics under Award No. DE-SC0020429. O.C. was supported by the U.S. National Science Foundation under Award PHY-1549132, the Center for Bright Beams. We would like to thank Prof. Richard Forbes for his constructive critique and valuable suggestions.

\bibliography{references}

\end{document}